\documentclass[journal]{IEEEtran}

\usepackage{graphicx}
\usepackage{epstopdf}

\usepackage{amssymb}
\usepackage{amsmath}
\usepackage{multirow}

\usepackage{siunitx}
\usepackage{times}
\usepackage[table,usenames,dvipsnames]{xcolor}
\usepackage{color}
\usepackage{balance}

\newcommand{\beqn}{\begin{equation}}
\newcommand{\eeqn}{\end{equation}}
\newcommand{\beqa}{\begin{eqnarray}}
\newcommand{\eeqa}{\end{eqnarray}}
\newcommand{\bear}{\begin{array}}
\newcommand{\eear}{\end{array}}
\newcommand{\bea}{\begin{align}}
\newcommand{\eea}{\end{align}}

\begin{document}

\title{Finding Structural Information about RF Power Amplifiers using an Orthogonal Non-Parametric Kernel Smoothing Estimator}

\author{Efrain~Zenteno,~\IEEEmembership{Student Member,~IEEE,}
        Zain~Ahmed~Khan,~\IEEEmembership{Student Member,~IEEE,}
        Magnus~Isaksson,~\IEEEmembership{Senior Member,~IEEE,}
        and~Peter~H\"andel,~\IEEEmembership{Senior Member,~IEEE}
\thanks{Copyright \textcopyright 2015 IEEE. Personal use of this material is permitted. However, permission to use this material for any other purposes must be obtained from the IEEE by sending a request to pubs-permissions@ieee.org.}
\thanks{E. Zenteno and Z.A Khan are with the Department of Electronics, Mathematics, and Natural Sciences, the University of G\"avle, SE 80176 G\"avle, Sweden, and also with the Department of Signal Processing, Royal Institute of Technology KTH, SE 10044 Stockholm, Sweden (e-mail: efnzeo@hig.se, zanahn@hig.se).
}
\thanks{M. Isaksson is with the Department of Electronics, Mathematics, and Natural Sciences, the University of G\"avle, SE 80176 G\"avle, Sweden (e-mail: min@hig.se).}
\thanks{P.  H\"andel is  with the  Department of Signal Processing, Royal Institute of Technology KTH, SE 10044 Stockholm, Sweden. (e-mail:  ph@kth.se)}}


\maketitle

\begin{abstract}
A non-parametric technique for modeling the behavior of power
amplifiers is presented. The proposed technique relies on the
principles of density estimation using the kernel method and is suited
for use in power amplifier modeling. 
 The proposed methodology  transforms the input domain into an
 orthogonal memory domain. In this domain, non-parametric static
 functions are discovered using the kernel estimator. These
 orthogonal, non-parametric functions can be fitted with any desired
 mathematical structure, thus facilitating its
 implementation. Furthermore, due to the orthogonality, the
 non-parametric functions can be analyzed and discarded individually,
 which simplifies pruning basis functions and provides a tradeoff between complexity and performance. 
The results show that the methodology can be employed to model power
amplifiers, therein yielding error performance similar to
state-of-the-art parametric models. Furthermore, a parameter-efficient
model structure with 6 coefficients was derived for a Doherty power amplifier,
therein significantly reducing the deployment's computational
complexity. Finally, the methodology can also be well exploited in digital linearization techniques. \end{abstract}


\IEEEpeerreviewmaketitle

\section{Introduction}
\IEEEPARstart{E}{nergy-efficient} power amplifiers (PAs) in wireless
networks usually behave in a nonlinear fashion, thereby producing
significant nonlinear distortions that degrade network
performance. This creates the need for suitable behavioral models for
PAs that provide simpler descriptions of nonlinear distortion
mechanisms and tools for the mitigation of these effects such as digital pre-distortion (DPD) techniques \cite{IsakssonWisellRonow2006}.

Historically, behavioral models for PAs have been derived using the
Volterra series \cite{Schetzen1980}. The disadvantage of the Volterra
series is that it involves a large number of parameters, which hinders
its practical implementation. Pruning Volterra series has been
actively studied to provide low-complexity and high-performance
behavioral models to mitigate PA nonlinear distortions
\cite{YouZhouChenZhou2014,Morgan2006,XinHong2013}. Although pruning
Volterra series has produced useful empirical models, these pruned
models are general structures for smaller classes of nonlinear
systems. This requires engineers to test different pruned model
structures and further select the nonlinearity order and memory depths
to meet certain performance requirements with a level of complexity
that depends on application constraints. Hence, for a specific PA,
trimming the pruned Volterra models may produce even lower complexity
with the desired error performance
\cite{YouZhouChenZhou2014,XinHong2013}. This raises the question as to
whether there may exist techniques to obtain structural knowledge of a
specific PA that in turn can be used to construct simpler model
structures with the required model error performance. This paper presents a technique of this class.

Trimmed model structures of reduced complexity can also be obtained
using sparse estimation techniques
\cite{ReinaAllegueCrespo2015,ZentenoAmin2014}. However, sparse
estimation techniques are usually computationally demanding and
require the choosing of an initial model to be reduced. On the one
hand, a general model is desired as an initial set that preserves the modeling properties. However,  this involves a large set, which increases the complexity of the technique. On the other hand, starting from a small class of model structures and  reducing complexity produces results that are dependent on this initial choice.

This paper presents a non-parametric method of discovering PA structural
information. Thus, it assumes no {\itshape{a priori}} model structure
for the PA. The proposed method considers static and dynamic
distortion effects and provides a tool for analyzing the PA transfer
function. In particular, the tool can be used to tailor parametric
models of simpler forms. Thus, the method effectively reduces the
computational complexity of the model. In PA modeling, other
non-parametric techniques use statistical functions such as cumulative
distribution functions (CDFs)
\cite{HuangHuangLeung2006,ZhiwenXipingCaron2013}  and  higher order
statistics  \cite{ZhiwenXinpingLeung2013}. However,
\cite{HuangHuangLeung2006,ZhiwenXipingCaron2013}, and
\cite{ZhiwenXinpingLeung2013} consider solely memoryless distortion
effects, and hence, they are ineffective at characterizing and compensating PA distortion caused by memory effects.

The proposed technique is based on non-parametric density estimation
\cite{Rosenblatt1956} referred to as the kernel smoothing estimator
method or simply the kernel method \cite{Georgiev1984}. Compared to
polynomial-based PA models, the kernel method can estimate nonlinear
functions of high nonlinearity order without numerical difficulties
\cite{KhanZenteno2014}. Furthermore, the kernel method uses window
averages, which are less computationally demanding than the matrix
(pseudo) inversions required in the least square methods. Finally, the
kernel estimator has strong statistical properties: asymptotic
convergence \cite{Epanechnikov1969}, optimal estimation in the square
error sense given a limited number of samples and robustness against
noise sources \cite{BaiYun2007}. All these properties make the kernel method a suitable candidate for modeling PAs.

The work reported in this paper reviews the modeling methodology
presented in \cite{Bai2008} and performs the adaptations necessary for
the PA measurement scenario. PAs are characterized by band-limited,
complex baseband equivalent signals, which make the method in \cite{Bai2008} unsuitable for PA modeling. However, with the adaptations proposed in \cite{JiangWilford2010} and our previous study \cite{KhanZenteno2014}, we obtain a methodology and method suitable for this application. 
In contrast to traditional PA modeling techniques, the work reported
here transforms the input sample domain into an orthogonal domain,
where the model structure is obtained using the kernel method. The
orthogonal domain simplifies the analysis of the PA transfer function;
allowing the addition or removal of basis functions provides a
tradeoff between complexity and performance. This result can be
transferred to the original sample domain, thereby reversing the
orthogonalization process (linear combination) and obtaining model
structures that are comparable in performance with the
state-of-the-art methods but with reduced computational requirements for deployment.

\section{PA Modeling}
\label{sec:background}

\subsection{PA model}

Let $u(n)$ and $y(n)$ denote the $n$-th complex-valued sample of the baseband signals corresponding to the input and output of a PA, respectively. The PA nonlinear transfer function is approximated by  \cite{Bai2008}
\begin{equation}
\begin{aligned}
y(n) &= \sum_{m_1} f_{m_1}(u(n-m_1)) + \\
 & + \sum_{m_1}\sum_{m_2} f_{m_1,m_2}(u(n-m_1),u(n-m_2)) + ... \\
 & + \sum_{m_1} ... \sum_{m_p} f_{m_1,m_2,\ldots,m_p}(u(n-m_1),...,u(n-m_p)),
\end{aligned}
\label{eq:NFIRsum}
\end{equation}
where $f_{m_1}(\cdot)$, $f_{m_1,m_2}(\cdot,\cdot)$ and $f_{m_1,m_2,\ldots,m_p}(\cdot,\ldots,\cdot)$ are nonlinear static functions whose domain dimensions are $1,2$ and $p$, respectively. The summations go up to $M$ subject to $0 \geq m_1 > m_2 > \ldots > m_p \geq M$, where $M$ is the maximum memory depth considered.

The Volterra series is a special form of \eqref{eq:NFIRsum}, which can be obtained when $f_{m_1}(\cdot)$, $f_{m_1,m_2}(\cdot,\cdot)$ and $f_{m_1,m_2,\ldots,m_p}(\cdot,\ldots,\cdot)$ are defined as the scaled product of their arguments. 
The static functions in \eqref{eq:NFIRsum} can represent high
nonlinearity orders of the Volterra series.  In particular, high
nonlinearity orders are coupled to different memory depths, which is
the cause of the rapid growth in the number of parameters in the Volterra
series. Despite the different features of \eqref{eq:NFIRsum} compared
to the Volterra series, both suffer from high
dimensionality. Considering $0 \geq m_1 > m_2 > \ldots > m_p \geq M$,
the system in \eqref{eq:NFIRsum}  has a total number of additive
functions of $\sum_{d=1}^{p}{M+1 \choose d} = \sum_{d=1}^{p}
\frac{(M+1)!}{d!\,(M+1-d)!}$, with $p$ being the highest dimension of
the functions in \eqref{eq:NFIRsum} and $!$ denoting the factorial
operator. The high dimensionality increases the computational
complexity of the identification and deployment of the models. Thus,
we analyze the relationship in \eqref{eq:NFIRsum} and study possible
simplifications to allow it to be suitable for PA modeling.

\subsection{Kernel method brief}
\begin{figure}[t]
	\centering
		\includegraphics[width=0.50\textwidth]{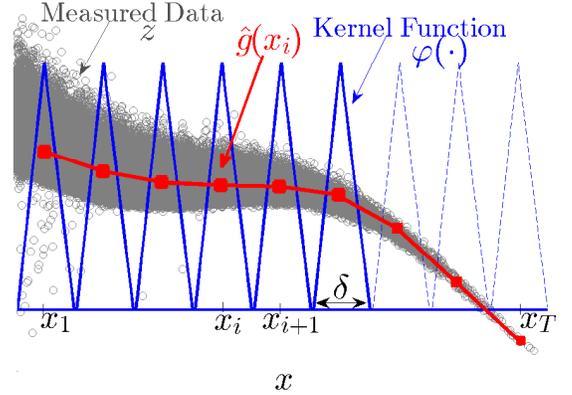}
	\caption{Illustration of the estimation of the function $g(\cdot)$ at fixed grid $x_i$ through a triangular kernel $\varphi(\cdot)$. The value $\hat{g}(x_i)$ is obtained as the weighted average of the output data $z$ through the kernel $\varphi(\cdot)$.}
	\label{fig:kernelExample}
\end{figure}

The kernel estimator as applied to the estimation of a static nonlinear input
output relation \cite{Georgiev1984} is briefly reviewed.  
Consider the set of real-valued input data $\{x(n)\}_{n=0}^{N-1}$
passed through an unknown, static nonlinear function $g(\cdot)$ and producing the output $\{z(n)\}_{n=0}^{N-1}$, that is, $z(n) = g(x(n))$ for $n=0,...,N-1$. Then, the static function $g(\cdot)$ can be estimated at a scalar point $x_i$ as illustrated in Fig. \ref{fig:kernelExample} by the kernel (window) average \cite{Georgiev1984}
\begin{equation}
\hat{g}(x_i) = \sum_{n=0}^{N-1} \frac{\varphi(\frac{x(n)-x_i}{\delta})}{\sum_{\ell=0}^{N-1}\varphi(\frac{x(\ell)-x_i}{\delta})} z(n) ,
\label{eq:functionID}
\end{equation}
where the grid of points $x_i$ for $i=1,...,T$ span the amplitude
support of $x(n)$ and $\varphi(\cdot)$ is the kernel with aperture
$\delta$.  Here, the triangular kernel is preferred because it is the
minimum mean square error estimator for a limited number of samples and
is robust against noise \cite{BaiYun2007}, that is,
\begin{equation}
\varphi(x) =
  \begin{cases}
   1-|x| & \text{if } |x| \leq 1 \\
   0     & \text{if } |x| > 1,
  \end{cases}
  \label{eq:kernelSVF}
\end{equation}
with $| \cdot |$ denoting the absolute value. Equation \eqref{eq:functionID} is evaluated only if the denominator is different from zero. In this paper, a linear interpolation between the two nearest neighbors is employed to compute $\hat{g}(\cdot)$ for an arbitrary input within the amplitude support of $x(n)$.

The kernel method is not directly suited to simulating PA
behavior. First, PA measurements show a significant correlation
between different samples of the input signal $u(n)$. The sample
correlation makes the output of the functions $f_{m_1}(\cdot)$,
$f_{m_1,m_2}(\cdot,\cdot)$ and
$f_{m_1,m_2,\ldots,m_p}(\cdot,\ldots,\cdot)$ jointly correlated. Thus,
their estimation needs to simultaneously account for all of them.

Second, the kernel method is intended for real-valued data. For
complex-valued data widely available within PA instrumentation, the
static functions in \eqref{eq:NFIRsum} are functions of complex-valued
inputs and outputs \cite{KhanZenteno2014}. This results in the
method having large computational and storage requirements.  In the
following, a method that addresses these drawbacks is outlined and discussed with simplifications (complexity reductions) of \eqref{eq:NFIRsum} suitable for PA modeling.

\subsection{Removing correlation by orthogonalization}
\label{sec:GS}
The PA input signal is band limited and digitized with oversampling, yielding $u(n)$. As a result of the oversampling, this discrete signal has a significant correlation between different samples. However, removing the correlation in $u(n)$ can be viewed as orthogonalizing it \cite{GramSchmidt2013}, which can be efficiently performed using the Gram-Schmidt (GS) process.

According to \eqref{eq:NFIRsum}, the signal set to be orthogonalized
lies in the space
\begin{equation}
{\mathcal{U}} = \left\{ u(n), \ldots, u(n-M) \right\}.
\label{eq:set2orth}
\end{equation}
The GS process yields an orthogonal set $\overline{\mathcal{U}} = \left\{ \bar{u}(n),\ldots, \bar{u}(n-M) \right\}$, where $\bar{u}(n) = u(n)$, followed by an iterative process for $k=1,...,M$,
\begin{equation}
\bar{u}(n-k) = u(n-k) - \sum_{\ell=0}^{k-1} P_{k,\ell} \bar{u}(n-\ell).
\label{eq:GS}
\end{equation}
The scalar $P_{k,\ell}$ is a projection of the signal $\bar{u}(n-\ell)$ over $u(n-k)$ defined by
\begin{equation}
P_{k,\ell}  = \sum_{n} u^{*}(n-k)\bar{u}(n-\ell),
\label{eq:proj}
\end{equation}
where $*$ denotes the complex conjugate operator. Note that the GS process involves a linear combination, and thus, it can be reversed without any loss of information. 
Assuming that $u(n)$ is a  wide-sense stationary stochastic process with
$k$-th auto-correlation lag denoted by $r_u(k)$, we note that the
projections can be {\itshape{a priori}} calculated using
$r_u(k)$. This leads to a computationally preferable method; e.g., a rectangular-shaped power spectral density of a Long-Term Evolution (LTE) signal yields a sinc-shaped auto-correlation function.

\subsection{Real-valued PA input signal}
\label{sec:URVD}
The distortion produced by PAs operating within wireless networks can be regarded as amplitude dependent \cite{Morgan2006,KhanZenteno2014}. Thus, considering solely the amplitude of the signals in the orthogonal input set $\overline{\mathcal{U}}$ yields the set
\begin{equation}
|\overline{\mathcal{U}}|=\left\{|\bar{u}(n)|,\ldots,|\bar{u}(n-M)|   \right\}, \label{eq:signaltransf}
\end{equation}
which will be the input to our kernel estimator, e.g., $x(n)=|\bar{u}(n)|$.
To compensate for the phase contribution, the output signal $y(n)$ is transformed as
\begin{equation} 
z(n) = y(n)e^{-j\angle{\bar{u}(n)}}.
\end{equation}
By applying the GS process to the input set ${\mathcal{U}}$ followed
by the real-value transformation, the system \eqref{eq:NFIRsum} becomes
\begin{equation}
\begin{aligned}
z(n) &= \sum_{m_1} g_{m_1}(x(n-m_1)) + \\
 & + \sum_{m_1}\sum_{m_2} g_{m_1,m_2}(x(n-m_1),x(n-m_2)) + ... \\
 & + \sum_{m_1} ... \sum_{m_p} g_{m_1,m_2...m_p}(x(n-m_1),...,x(n-m_p)),
\end{aligned}
\label{eq:NFIRsumORT}
\end{equation}
with complex-valued $g_{m_1}(\cdot)$, $g_{m_1,m_2}(\cdot,\cdot)$ and
$g_{m_1,m_2,\ldots,m_p}(\cdot,\ldots,\cdot)$ as the orthogonal
counterparts of the functions in \eqref{eq:NFIRsum} but with
real-valued arguments. This reduces the estimation dimension required
in the kernel method. The system \eqref{eq:NFIRsumORT} has similar
features  as \eqref{eq:NFIRsum} for modeling nonlinear behavior.
However, in contrast to \eqref{eq:NFIRsum},  it has orthogonal basis
functions. Thus, their estimation can be performed individually, and each
basis contribution can be separately analyzed.

\subsection{Complexity reduction of \eqref{eq:NFIRsum}}
\label{sec:ComRed}

Despite using real-valued input signals, the complexity of the
estimation of a multi-variable function in \eqref{eq:NFIRsumORT}
remains high; e.g., a $p$-th variable function estimated at $T$
points for each variable gives a total of $T^p$ estimation
points. Thus, the memory requirements and data manipulation increase
exponentially with $p$, leading to the well-known {\itshape{curse of
    dimensionality}} problem. In an attempt to alleviate this, a $p$-variable function $g_{m_1,...,m_p}(\cdot,\ldots,\cdot)$ is approximated as a sum of single-variable functions:
\begin{equation}
\begin{aligned}
g_{m_1,...,m_p} (& x(n-m_1),...,x(n-m_p) )  \approx \\
& \sum_{k=1}^{p} h_{m_k} ( x(n-m_k)\prod_{\begin{subarray}{c} d=1 \\ d \neq k \end{subarray}}^{p} x(n-m_d) ).
\end{aligned}
\label{eq:RedCompl}
\end{equation}
Thus, the single-variable functions $h_{m_k}(\cdot)$ can be estimated using \eqref{eq:functionID}. In PA modeling, \eqref{eq:RedCompl} has been motivated from a physical \cite{CunhaLimaPedro2010} and signal processing perspective \cite{JiangWilford2010}.
Note that the new single-variable $x(n-m_k)\prod_{\begin{subarray}{c} d=1 \\ d \neq k \end{subarray}}^{p} x(n-m_d) $ can similarly be  considered in the non-orthogonal domain ${\mathcal{U}}$ by augmenting it as
\begin{equation}
{\mathcal{U}}' = \left[ {\mathcal{U}}, \\ \{ u(n-m_k)\prod_{\begin{subarray}{c} d=1 \\ d \neq k \end{subarray}}^{p}| u(n-m_d)| \}_{m_k=0}^M \right]  
 \label{eq:set2orthE}
\end{equation} 
for $p=2,\ldots,M$. The orthogonalization of the data set ${\mathcal{U}}'$ is performed using the GS procedure.

\subsection{Summary and implementation}

Consider the data set of complex-valued input and output measurements
$\{u(n)\}$ and $\{y(n)\}$, respectively. The non-parametric modeling
approach begins by creating the input space $\mathcal{U}'$ as
indicated by \eqref{eq:set2orthE} for the chosen maximum memory depth
$M$. The implementation of the method can proceed by storing
$\mathcal{U}'$ as a matrix whose columns are the basis in
\eqref{eq:set2orthE}. This matrix is column-wise orthogonalized using
the GS process, therein yielding an orthogonal matrix. Only magnitude entries of the orthogonal matrix are retained according to \eqref{eq:signaltransf}, rendering a magnitude matrix. Finally, each column of the magnitude matrix is used as a domain to estimate single-variable functions with the kernel method in \eqref{eq:functionID}.
The results show that significant contribution to the model output
continues to originate from a few single-variable functions. Thus, due to the orthogonality, the non-contributing functions can be eliminated from the model structure (remove the corresponding columns) while retaining the obtained model performance.

\section{Experimental}
\label{sec:experimental}

\subsection{Measurement setup}
The measurement setup  includes a vector signal generator R\&S SMU
200A that is used to excite the PA. The PA output is measured using a
wideband down converter and a high-performance analog-to-digital
converter (ADC) with 14-bit resolution operated with a 400 MHz
sampling rate. The amplifier being tested is the MRF8S21120HS Doherty
amplifier with 14 dB linear gain, an operation frequency in the
2.1-2.2 GHz band, and a rating of 46 dBm output power operated at approximately 3 dB of compression. 


Two independent excitation signals  with bandwidths of 12 and 24 MHz
are generated. These excitations are noise-like signals with
peak-to-average power ratios of 11.2 and 11.4 dB, respectively. The
excitations were  created in a PC using $10^5$ complex-valued samples
uploaded to the generator and up-converted to 2.14 GHz to excite the PA.  The measurements consist of $10^5$ complex-valued samples for the input and the output of the amplifier with post-processing time and phase delay compensation \cite{IsakssonWisellRonow2006}. The non-parametric structure is obtained using 10$\%$ (estimation phase) of the measured data, and the remaining 90$\%$ (validation phase) is used to evaluate the modeling error.

\subsection{Results}

\subsubsection{User-defined parameters}
\begin{figure}[t]
	\centering
		\includegraphics[width=0.50\textwidth]{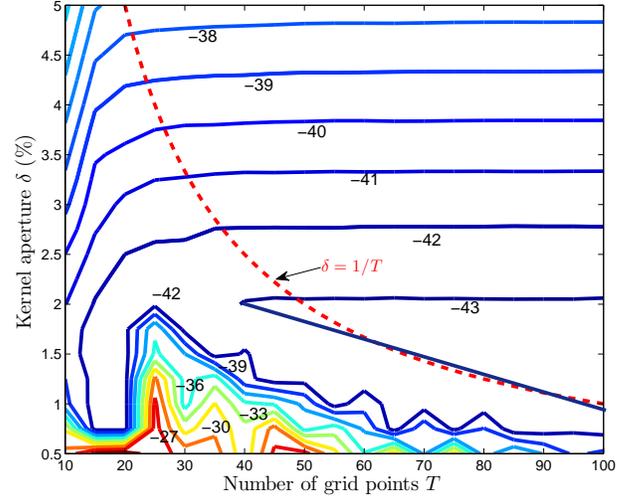}
	\caption{NMSE (in dB) as a function of the number of grid
          points $T$ and the kernel aperture $\delta$ (as a percentage of the span of the input signal).}
	\label{fig:deltavsT}
\end{figure}

In the proposed method, the number of grid points $T$ and the kernel aperture $\delta$ are user-defined parameters.  Fig. \ref{fig:deltavsT} shows the normalized mean square error (NMSE) contours over both $\delta$ and $T$ in a linearly spaced grid.  The kernel aperture $\delta$ is shown as a percentage of the span of the input signal.

The number of grid points $T$ sets the resolution of the static
function and the kernel aperture $\delta$ sets the size of the input
neighborhood to perform the average (estimation) (cf.
Fig. \ref{fig:kernelExample}).  Thus, a large value of $T$ and a small
value of
$\delta$ are desired to produce an accurate estimation. However, for a
fixed number $N$ of measurement samples, decreasing $\delta$ may
degrade the performance because the number of measurement samples in
each kernel function decreases, and hence, its average (estimate) has increased variance (less reliability).  This is the reason for the loss in performance for small values of $\delta$ in Fig. \ref{fig:deltavsT}.
Because $T$ is the number of entries to be stored, $T$ can be chosen
based on the available memory resources, and as a rule of thumb, the
kernel aperture can be set as $\delta = 1/T$ to avoid the performance
degrading effects, cf. Fig. \ref{fig:deltavsT}. The choice  $\delta =
1/T$ has the advantage of efficiently using all training data for estimating the non-parametric model.

\subsubsection{Modeling performance}

Using $T=70$ and  $\delta=1/70$, a non-parametric model of the PA is obtained for the two input signals under consideration. Table~\ref{tab:NmseAcepr2} shows the individual contributions of the basis functions to the NMSE and to the adjacent channel error power ratio (ACEPR) \cite{IsakssonWisellRonow2006} for these two signals. 
As observed in Table~\ref{tab:NmseAcepr2}, the function $\hat{g}_0(x)$ contributes  -38.3 and -32.5 dB of the total NMSE for the 12 and 24 MHz signals, respectively. This function is the largest model contributor because it captures linear and nonlinear static effects.

\begin{table}[t]
\renewcommand{\arraystretch}{1.3}
\caption{Individual contributions of the functions for two different signal bandwidths}
\centering
\begin{tabular}{ c|c|c|c|c }
    \hline\hline
    \multicolumn{1}  { c| }{ \multirow{2}{*}{Basis} } & \multicolumn{2}{ c| }{ 12 MHz} & \multicolumn{2}{ c }{ 24 MHz} \\    	\cline{2-5}
					 		& NMSE	& ACEPR 	& NMSE 	& ACEPR \\  
							& [dB] & [dB] & [dB] & [dB] \\ \cline{1-5}	
	$\hat{g}_0(\cdot)$ 		& \textcolor{blue}{\bfseries{-38.3}} & -54.9 	& \textcolor{blue}{\bfseries{-32.5}} 	& -50.8 			\\
	$\hat{g}_1(\cdot)$ 		& \textcolor{blue}{\bfseries{-5.0}}	& 0.0 		& \textcolor{blue}{\bfseries{-6.7}} 	& -0.4				\\
	$\hat{g}_2(\cdot)$ 		& -0.3 	& 0.0 		& \textcolor{blue}{\bfseries{-2.3}} 	& 0.3	 			\\
	$\hat{g}_3(\cdot)$ 		& -0.0 	& 0.0 		& -0.4 	& 0.0	 			\\   \hline
	$\hat{g}_{0,1}(\cdot)$ 	& -0.1	&  -0.3	&  \textcolor{blue}{\bfseries{-0.5}}  & 	-0.2 			\\
	$\hat{g}_{0,2}(\cdot)$ 	& -0.1  &  0.0	&  -0.1	 &  0.1		 		\\
	$\hat{g}_{0,3}(\cdot)$ 	& -0.0	&  0.0	&  -0.0	 &  0.0				\\
	$\hat{g}_{1,2}(\cdot)$ 	& -0.0	&  0.0	&  0.0		& 0.0 				\\
	$\hat{g}_{1,3}(\cdot)$ 	&  0.0	&  0.0	&  0.0		& 0.0 				\\
	$\hat{g}_{2,3}(\cdot)$ 	&  0.0	&  0.0	&  0.0		& 0.0 				\\  \hline
	$\hat{g}_{0,1,2}(\cdot)$ 	& -0.1 	&  0.0	&  -0.1	 & -0.0					\\
	$\hat{g}_{0,1,3}(\cdot)$ 	&  0.0	&  0.0	&  	0.0	 & 0.0 				\\
  $\hat{g}_{0,2,3}(\cdot)$ 	&  0.0	&  0.0	&  	0.0	 & 0.0 				\\
	$\hat{g}_{1,2,3}(\cdot)$ 	&  0.0	&  0.0	&  	0.0	 & 0.0 				\\    \hline
	Total       &  -43.9 &  -55.2	&  	-42.6	 & -51.0 		\\  	\hline\hline
\end{tabular}
\label{tab:NmseAcepr2}
\end{table}

For the 12 MHz signal, the NMSE is dominated by the contributions of
$\hat{g}_0(x)$ and $\hat{g}_1(x)$, which provide a combined NMSE of
-43.3 dB. However, in the 24 MHz input signal, the contribution to the
NMSE from the function $\hat{g}_2(x)$ increases from -0.3 to -2.3 dB,
thereby revealing the impact of memory effects caused by the increase
in signal bandwidth. Note that the contributions from the 2- and
3-variable functions is negligible in the 12 MHz case, and
$\hat{g}_{0,1}(\cdot)$ significantly increases its contribution when increasing the signal bandwidth from 12 to 24 MHz from -0.1 to -0.5 dB.  In terms of the ACEPR,  only $\hat{g}_0(x)$  provides a significant contribution for both signal bandwidths.  
The static nonlinear distortion is modeled by $\hat{g}_0(\cdot)$, and the linear and nonlinear dynamics are described by $\hat{g}_1(\cdot)$ and $\hat{g}_2(\cdot)$. Because these functions are arbitrary,
the non-parametric structure can model memory effects {\itshape{coupled}} to strong nonlinearities, which are one of the causes of poor behavior in polynomial-based model methodologies.

\begin{figure}[t]
\centering
\includegraphics[width=0.50\textwidth]{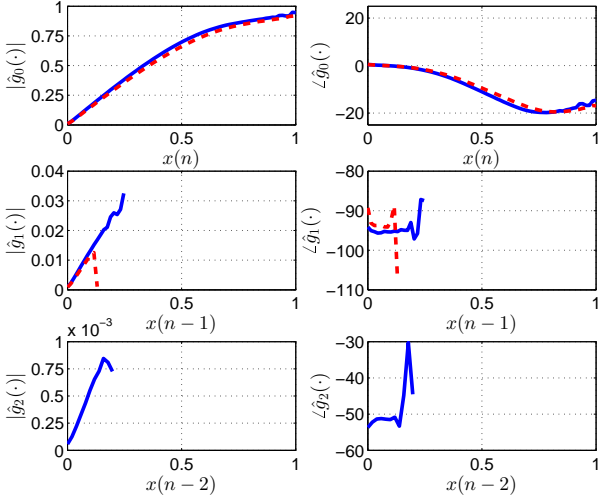}
\caption{Amplitude and phase (in degrees) of the estimated
  single-variable functions $\hat{g}_0(\cdot)$, $\hat{g}_1(\cdot)$,
  and $\hat{g}_2(\cdot)$ for the 24 MHz signal (solid - blue) and
  $\hat{g}_0(\cdot)$ and $\hat{g}_1(\cdot)$ for the 12 MHz (dashed - red)
  signal.   The function $\hat{g}_2(\cdot)$ is not presented for the 12 MHz signal because its contribution is negligible.} 
\label{fig:Kernels}
\end{figure}

The functions contributing more than -2 dB to the NMSE are shown in Fig. \ref{fig:Kernels} for both signal bandwidths. Despite the two signals being independently created and having different bandwidths, the estimated functions are similar to each other, which suggests that the method obtains structural information about the modeled PA.

\begin{figure}[t]
\centering
\includegraphics[width=0.50\textwidth]{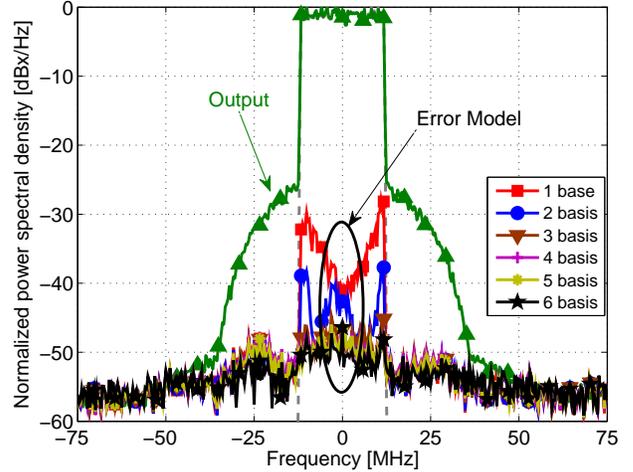}
\caption{Power spectral density of the output and error model obtained
  by sequentially including the first 6 basis functions indicated in Table~\ref{tab:NmseAcepr2} in the nonparametric model for the 24 MHz signal.}
\label{fig:EPSD24}
\end{figure}

The power spectral density (PSD) of the input, output and model error evolution are plotted in Fig.~\ref{fig:EPSD24} for the 24 MHz bandwidth signal. The model was updated sequentially to include the first six  basis functions of Table~\ref{tab:NmseAcepr2}. The in-band error spectrum decreases with the addition of basis functions, and the out-of-band error is suppressed by the use of $\hat{g}_0(x)$. These two observations are in accordance with Table~\ref{tab:NmseAcepr2}.

An advantage of the proposed method is that we can utilize the
estimated basis functions (cf. Fig. \ref{fig:Kernels}) to build a
parametric model of the PA. These parametric models can be of any
form; they can be chosen to ease implementation and identification or to maximize performance. We seek a parameter efficient representation using a polynomial family as an example.
Thus, the function $\hat{g}_0(\cdot)$ is modeled with a seventh-order
polynomial, and $\hat{g}_1(\cdot)$ and $\hat{g}_2(\cdot)$ are modeled with linear polynomials (cf. Fig. \ref{fig:Kernels}), that is, 
\begin{equation}
z(n) = \sum_{p=1}^{4} \gamma_p \bar{u}(n) |\bar{u}(n)|^{2(p-1)} +  \gamma_{5} \bar{u}(n-1) + \gamma_{6} \bar{u}(n-2).
\label{eq:6parOr}
\end{equation}
Due to the orthogonality, the remainder of the functions in Table~\ref{tab:NmseAcepr2} can be discarded without affecting the model performance.
Furthermore, by replacing the orthogonal variables for the linear combination of their non-orthogonal counterparts given in the GS process (Section \ref{sec:GS}), the parametric description of this model becomes 
\begin{equation}
y(n) = \sum_{p=1}^{4} \alpha_p u(n) |u(n)|^{2(p-1)} +  \alpha_{5} u(n-1) + \alpha_{6} u(n-2).
\label{eq:6par}
\end{equation}
This 6-parameter model $[\alpha_1, \ldots, \alpha_6]$ can be
identified using linear regression techniques, which are commonly used in PA modeling \cite{Morgan2006}.

\begin{figure}[t]
\centering
\includegraphics[width=0.50\textwidth]{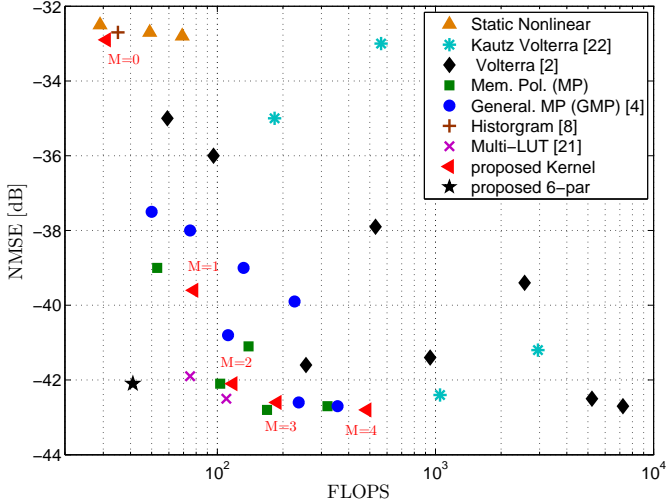}
\caption{NMSE of the 24 MHz signal versus the number of FLOPS incurred
  when using the feed-forward model for several modeling techniques. The kernel method uses $\delta=1/T,\ T = 70$ with $M$ memory depth. Different NMSE values are obtained by changing the settings in the methods.}
\label{fig:comparison}
\end{figure}

Fig. \ref{fig:comparison} shows the NMSE performance versus the
complexity incurred when using the feed-forward model. The complexity
is measured in floating point operations (FLOPs)
\cite{TehraniHaiying2010}. Fig. \ref{fig:comparison} compares the
proposed method with several parametric models, such as the static
nonlinear, memory polynomial, generalized memory polynomial
\cite{Morgan2006}, Multi-LUT \cite{GilabertCesariMontoro2008},
Volterra \cite{Schetzen1980} and Kautz-Volterra \cite{IsakssonRonnow2007} models,
and non-parametric models, such as the Histogram model \cite{HuangHuangLeung2006}.
Different points correspond to different model settings (nonlinearity order and memory depth) being tested. 

Although, in general, it is possible to trade reduced  NMSE for increased complexity, these settings need to be chosen with care to obtain optimal performance for the level of complexity incurred, as observed by the performance dispersion in Fig. \ref{fig:comparison}.
Moreover, some model settings with increased model complexity degrade
the NMSE performance, as observed in Fig. \ref{fig:comparison}, which
is due to an unsuitable model being chosen. For example, a static
model of high nonlinearity order may have large complexity but remains
unable to model dynamic effects, thereby providing limited
performance. Similar arguments can be made for memory
depths. Moreover, these detrimental effects have been discussed in
previous studies \cite{TehraniHaiying2010}. The proposed kernel method
has good performance/complexity compared to state-of-the-art
parametric models. Finally, the proposed method was used to construct
a parameter-efficient structure (6-parameter model), which provides
the best NMSE for its reduced level of complexity because it was
specifically tailored for the PA.

\section{Digital Pre-distortion (DPD)}
\label{sec:DPD}
The proposed method is tested as a pre-distorter compensating for
nonlinear distortions at the PA output. The non-parametric structure
is obtained using an inverse learning architecture, in which input and
output are interchanged \cite{EunPowers1997}. To increase efficiency,
a clipping technique \cite{JiangWu2008} has been applied to the 24 MHz
input signal, therein reducing its PAPR from 11.4 to 8.8 dB. However, care must be exercised because clipping techniques introduce in-band and out-of-band errors.

\begin{figure}[t]
\centering
\includegraphics[width=0.50\textwidth]{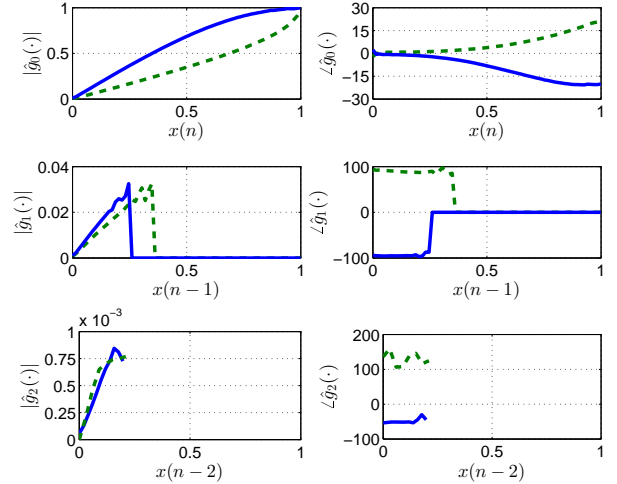}
\caption{Amplitude and phase (in degrees) of the estimated functions of the feed-forward model in solid blue and inverse (DPD) model in dashed green. The inverse model was estimated using the inverse learning architecture \cite{EunPowers1997}.}
\label{fig:DPDresult}
\end{figure}

From \eqref{eq:NFIRsumORT}, the function $g_{0}(\cdot)$ of the DPD
model has to be the inverse of the same function in the feed-forward
model. For the remainder of the functions in the DPD model, they have
to be the negative of their counterparts in the feed-forward model
(same amplitude but with the phase shifted by 180 degrees). This is depicted in Fig. \ref{fig:DPDresult}, where the feed-forward and inverse (DPD) estimated non-parametric functions are plotted.

The PA operates at -25 dB of NMSE and -36 dB of ACPR, respectively,
without DPD. The pre-distorted PA with the outlined method produces an
NMSE and ACPR of -42 and -49.5 dB, respectively, which shows its
effectiveness in compensating nonlinear distortion.

\section{Conclusions}
\label{sec:conc}

A non-parametric method of modeling RF power amplifiers is
presented. The method does not assume an {\itshape{a priori}} model
structure of the PA. Thus, basis functions that describe its behavior
are estimated during the identification process, leading to the
development of tailored parametric models. These tailored models can
be fitted with any desired structure, which eases its implementation. In
particular, parameter-efficient models with small errors can be
obtained, thereby reducing the implementation and deployment computational costs.

The presented method is based on the kernel estimator, which solely
performs sample averages and hence does not suffer from numerical
instabilities. Furthermore, adaptive schemes can be made using running
averages, which require low computational resources and feature
real-time implementations. The proposed methodology can lead to the low computational resource implementation of look-up tables (LUTs) for adaptive digital pre-distortion (linearization).


\bibliographystyle{IEEEtran}
\bibliography{refe}

%
\begin{IEEEbiographynophoto}{Efrain Zenteno}
(S'10) received the B.S. degree from San Agustin University, Arequipa, Peru, in 2004. He is currently pursuing the Ph.D. degree at the Department of Signal Processing, Royal Institute of Technology KTH, Stockholm, Sweden. During 2005, he was a field engineer at the electric company SEAL. He received the M.Sc.degree in electronics/telecommunications engineering from the University of G\"avle, G\"avle, Sweden, in 2008.
From 2008 to 2010, he was with the Program of Telecommunications Engineering, Universidad Católica San Pablo, Arequipa, Peru. He is currently with the Department of Electronics, Mathematics, and Natural Sciences, University of G\"avle, and with the Department of Signal Processing, Royal Institute of Technology KTH. His main interests are instrumentation, measurements, and signal processing algorithms for communications.
\end{IEEEbiographynophoto}

\begin{IEEEbiographynophoto}{Zain Ahmed Khan}
(S'13) received his B.S. degree in electronics engineering from Ghulam Ishaq Khan Institute, Pakistan, in 2005 and his M.S. degree in communication and signal processing from Technical University Ilmenau, Germany in September 2013. Currently, he is a PhD student at the Department of Signal Processing, Royal Institute of Technology KTH, Stockholm, Sweden and is working at the Department of Electronics, Mathematics, and Natural Sciences, University of G\"avle. He started his PhD in December 2013 and is working on the behavioral modeling and digital predistortion of RF power amplifiers with particular focus on MIMO transmitters. 
\end{IEEEbiographynophoto}

\begin{IEEEbiographynophoto}{Magnus Isaksson}
(S'98-\-M'07-\-SM'12) received the M.Sc. degree in microwave engineering from the University of G\"avle, G\"avle, Sweden, in 2000, the Licentiate degree from Uppsala University, Uppsala, Sweden, in 2006, and the Ph.D. degree from the Royal Institute of Technology, Stockholm, Sweden, in 2007. In 2012, he was appointed Docent in Telecommunications in the Royal Institute of Technology (KTH), Stockholm, Sweden. During 1989-1999, he was with the Televerket, Sweden, working on communication products. Since 1999, he is with the Department of Electronics, Mathematics, and Natural Sciences at the University of G\"avle, G\"avle, Sweden, where he is currently a professor of electronics and department head. His research activities are in signal processing algorithms for radio-frequency measurements and characterization, modeling, and compensation of nonlinear microwave devices and systems. Dr. Isaksson is the author or co-author of many published peer-review journal articles, books, and conference proceedings in the area, and is currently Head of research within the fields of mathematics, and natural sciences at the University of G\"avle, G\"avle, Sweden.
\end{IEEEbiographynophoto}


\begin{IEEEbiographynophoto}{Peter H\"andel}
(S'88-\-M'94-\-SM'98) received the Ph.D. degree from Uppsala University, Uppsala,
Sweden, in 1993. From 1987 to 1993, he was with Uppsala University. From 1993 to 1997, he was with Ericsson AB, Kista, Sweden. From 1996 to 1997, he was a Visiting Scholar with the Tampere University of Technology, Tampere, Finland. Since 1997, he has been with the KTH Royal Institute of Technology, Stockholm, Sweden, where he is currently a Professor of Signal Processing and Head of the Department of Signal Processing. From 2000 to 2006, he held an adjunct position at the Swedish Defence Research Agency. He has been a Guest Professor at the Indian Institute of Science (IISc), Bangalore, India, and at the University of G\"avle, Sweden. He is a co-founder of Movelo AB. Dr. H\"andel has served as an associate editor for the IEEE TRANSACTIONS ON SIGNAL PROCESSING. He was a recipient of a Best Survey Paper Award by the IEEE Intelligent Transportation Systems Society in 2013.
\end{IEEEbiographynophoto}

\balance

\end{document}